\documentclass[preprint]{aastex63}

\newcommand{\fe}{\ion{Fe}{1}}
\newcommand{\fea}{\ion{Fe}{1} $\lambda$6569}
\newcommand{\ha}{H$\alpha$}
\newcommand{\alfven}{Alfv\'{e}n}

\usepackage{amsmath}
\usepackage{color}
\usepackage{multirow}

\begin{document}

\title{Spectral observations and modeling of a solar white-light flare observed by CHASE}

\author[0000-0003-0057-6766]{De-Chao Song}
\affiliation{Key Laboratory of Dark Matter and Space Astronomy, Purple Mountain Observatory, CAS, Nanjing 210023, People's Republic of China}
\affiliation{School of Astronomy and Space Science, University of Science and Technology of China, Hefei 230026, People's Republic of China}

\author[0000-0002-1068-4835]{Jun Tian}
\affiliation{Key Laboratory of Dark Matter and Space Astronomy, Purple Mountain Observatory, CAS, Nanjing 210023, People's Republic of China}
\affiliation{School of Astronomy and Space Science, University of Science and Technology of China, Hefei 230026, People's Republic of China}

\author[0000-0002-8258-4892]{Y. Li}
\affiliation{Key Laboratory of Dark Matter and Space Astronomy, Purple Mountain Observatory, CAS, Nanjing 210023, People's Republic of China}
\affiliation{School of Astronomy and Space Science, University of Science and Technology of China, Hefei 230026, People's Republic of China}

\author[0000-0002-4978-4972]{M. D. Ding}
\affiliation{School of Astronomy and Space Science, Nanjing University, Nanjing 210023, People's Republic of China}
\affiliation{Key Laboratory for Modern Astronomy and Astrophysics (Nanjing University), Ministry of Education, Nanjing 210023, People's Republic of China}

\author[0000-0002-4241-9921]{Yang Su}
\affiliation{Key Laboratory of Dark Matter and Space Astronomy, Purple Mountain Observatory, CAS, Nanjing 210023, People's Republic of China}
\affiliation{School of Astronomy and Space Science, University of Science and Technology of China, Hefei 230026, People's Republic of China}

\author[0000-0003-2872-2614]{Sijie Yu}
\affiliation{Center for Solar-Terrestrial Research, New Jersey Institute of Technology, 323 Martin Luther King Jr. Blvd., Newark, NJ 07102-1982, USA}

\author[0000-0002-8002-7785]{Jie Hong}
\affiliation{School of Astronomy and Space Science, Nanjing University, Nanjing 210023, People's Republic of China}
\affiliation{Key Laboratory for Modern Astronomy and Astrophysics (Nanjing University), Ministry of Education, Nanjing 210023, People's Republic of China}

\author[0000-0002-1190-0173]{Ye Qiu}
\affiliation{School of Astronomy and Space Science, Nanjing University, Nanjing 210023, People's Republic of China}
\affiliation{Key Laboratory for Modern Astronomy and Astrophysics (Nanjing University), Ministry of Education, Nanjing 210023, People's Republic of China}

\author[0000-0002-7544-6926]{Shihao Rao}
\affiliation{School of Astronomy and Space Science, Nanjing University, Nanjing 210023, People's Republic of China}
\affiliation{Key Laboratory for Modern Astronomy and Astrophysics (Nanjing University), Ministry of Education, Nanjing 210023, People's Republic of China}

\author[0000-0002-3657-3172]{Xiaofeng Liu}
\affiliation{Key Laboratory of Dark Matter and Space Astronomy, Purple Mountain Observatory, CAS, Nanjing 210023, People's Republic of China}
\affiliation{School of Astronomy and Space Science, University of Science and Technology of China, Hefei 230026, People's Republic of China}

\author[0000-0001-7540-9335]{Qiao Li}
\affiliation{Key Laboratory of Dark Matter and Space Astronomy, Purple Mountain Observatory, CAS, Nanjing 210023, People's Republic of China}
\affiliation{School of Astronomy and Space Science, University of Science and Technology of China, Hefei 230026, People's Republic of China}

\author[0000-0002-1810-6706]{Xingyao Chen}
\affiliation{School of Physics \& Astronomy, University of Glasgow, Glasgow G12 8QQ, UK}

\author[0000-0001-7693-4908]{Chuan Li}
\affiliation{School of Astronomy and Space Science, Nanjing University, Nanjing 210023, People's Republic of China}
\affiliation{Key Laboratory for Modern Astronomy and Astrophysics (Nanjing University), Ministry of Education, Nanjing 210023, People's Republic of China}

\author{Cheng Fang}
\affiliation{School of Astronomy and Space Science, Nanjing University, Nanjing 210023, People's Republic of China}
\affiliation{Key Laboratory for Modern Astronomy and Astrophysics (Nanjing University), Ministry of Education, Nanjing 210023, People's Republic of China}

\correspondingauthor{Y. Li and M. D. Ding}
\email{yingli@pmo.ac.cn, dmd@nju.edu.cn}
 
\begin{abstract}
The heating mechanisms of solar white-light flares remain unclear. We present an X1.0 white-light flare on 2022 October 2 (SOL2022-10-02T20:25) observed by the Chinese \ha\ Solar Explorer (CHASE) that provides two-dimensional spectra in the visible light for the full solar disk with a seeing-free condition. The flare shows a prominent enhancement of $\sim$40\% in the photospheric \fe\ line at 6569.2 \AA, and the nearby continuum also exhibits a maximum enhancement of $\sim$40\%. For the continuum near the \fe\ line at 6173 \AA\ from the Helioseismic and Magnetic Imager (HMI) on board the Solar Dynamics Observatory (SDO), it is enhanced up to $\sim$20\%. At the white-light kernels, the \fe\ line at 6569.2 \AA\ has a symmetric Gaussian profile that is still in absorption and the H$\alpha$ line at 6562.8 \AA\ displays a very broad emission profile with a central reversal plus a red or blue asymmetry. The white-light kernels are co-spatial with the microwave footpoint sources observed by the Expanded Owens Valley Solar Array (EOVSA) and the time profile of the white-light emission matches that of the hard X-ray emission above 30 keV from the Gamma-ray Burst Monitor (GBM) on Fermi. These facts indicate that the white-light emission is qualitatively related to a nonthermal electron beam. We also perform a radiative hydrodynamic simulation with the electron beam parameters constrained by the hard X-ray observations from Fermi/GBM. The result reveals that the white-light enhancement cannot be well explained by a pure electron-beam heating together with its induced radiative backwarming but may need additional heating sources such as \alfven\ waves. 
\end{abstract}



\section{Introduction} 
\label{sec:intro}

White-light flares (WLFs) belong to a relatively rare type of flares characterized by a sudden increase in the visible continuum. The first solar flare ever observed, the so-called Carrington event \citep[][]{Carrington1859}, was a typical WLF. WLFs have also been observed on other stars \citep[e.g.,][]{Pugh2016}. Solar WLFs can be divided into two types \citep[e.g.,][]{Machado1986}. Type I WLFs have a well-established correlation between the peak time of continuum emission and that of hard X-ray (HXR) and microwave emissions, which also show an evident Balmer jump in the continuous spectrum \citep[e.g.,][]{Fang1995AAS}. By contrast, type II WLFs lack those features \citep[e.g.,][]{Ding1999,Prochazka18}. It should be noted that some WLFs show observational characteristics in between these two types, highlighting a complexity of the heating mechanisms underlying WLFs \citep[e.g.,][]{Hiei1982,hao17nc}.

In recent decades, various heating mechanisms have been proposed for WLFs based on observations and simulations, including nonthermal electron beam bombardment \citep[e.g.,][]{Hudson1992,Krucker2015,Watanabe2020}, proton beam bombardment \citep[e.g.,][]{Machado1978,Henoux1993,Prochazka19}, radiative backwarming \citep[e.g.,][]{Metcalf1990,Ding2003,Song2018}, chromospheric condensation \citep[e.g.,][]{Gan1992,Kowalski2015}, \alfven\ wave dissipation \citep[e.g.,][]{Emslie1982,Fletcher2008}, and soft X-ray (SXR)/EUV irradiation \citep[e.g.,][]{Henoux&Nakagawa1977A&A,Poland1988}. Nevertheless, the heating mechanisms of WLFs have not been fully understood. In some events, it is likely that the enhancement of white-light emission is caused by a joint effect of multiple mechanisms as mentioned above \citep[e.g.,][]{Machado1978,Xu2010}. 

Besides the continuum, spectral lines like the \fe\ and \ha\ lines have been used to diagnose the heating mechanisms of WLFs \citep[e.g.,][]{Lin1996,Baranovsky1997,Yang2021,hong22}. \cite{Babin1992} reported that in an X12 WLF, several \fe\ lines in the wavelength range of 6517--6598 \AA\ lack an absorption in the profiles. \cite{Jurcak2018} observed emission profiles in the \fe\ $\lambda$6301 and $\lambda$6302 lines in an X9.3 WLF and judged that the continuum enhancement mainly originates from the heated chromosphere rather than the photosphere. In an M2.0 flare, the \fe\ $\lambda$6173 line was found to remain in absorption and become blueshifted \citep{Martinez2011}. Based on radiative hydrodynamic simulations in a flaring atmosphere, \cite{hong18} found that the enhancement of the \fe\ $\lambda$6173 line can be caused by both electron beam heating in the lower chromosphere and radiative backwarming in the photosphere. Regarding the \ha\ $\lambda$6563 line, it is much more complicated. For example, it can be double-peaked with usually a red asymmetry in WLFs \citep{Babin1992,Zhou1997}. 
 
In this Letter, we present an X1.0 WLF on 2022 October 2 well observed by the Chinese \ha\ Solar Explorer \citep[CHASE;][]{lich19,lich22}, which provides spectral observations in the \fea\ and \ha\ lines simultaneously with a seeing-free condition. This is the first X-class flare captured by CHASE. It shows a prominent enhancement of $\sim$40\% in the \fe\ line and the nearby continuum can also be enhanced up to 40\%. The white-light brightenings are related to HXR and microwave emissions spatially or temporally. A radiative hydrodynamic simulation with the parameters constrained by HXR observations reveals that such a prominent enhancement cannot be explained by a pure electron-beam heating together with its induced radiative backwarming but needs additional heating sources.


\section{Observations and data reduction}

The X1.0 flare under study was observed by multiple space- and ground-based telescopes. CHASE captured the main phase of the flare and acquired the two-dimensional spectra of \fe\ $\lambda$6569 (6567.5--6569.8 \AA) and \ha\ $\lambda$6563 (6559.4--6565.2 \AA) simultaneously for the full disk via a raster scanning mode. The spatial, temporal, and spectral resolutions of CHASE imaging spectral data in a two-binning mode here are $\sim$1\arcsec\ pixel$^{-1}$, $\sim$1 minute, and 0.05 \AA\ pixel$^{-1}$, respectively. The data have been processed with corrections for dark field, flat field, and slit image curvature, as well as wavelength calibration \citep{qiuy22}. When measuring the relative enhancements of the continuum, \fe\ line, and \ha\ line intensities, we calculate the standard deviation of intensities in difference images for a relatively quiet region as their uncertainties. The \fe\ and \ha\ lines are analyzed by adopting a single Gaussian fitting and a bisector method, respectively. In order to quantitatively measure the Doppler velocity, we need an accurate reference wavelength that is obtained by averaging the line profiles over the quiet region. This yields velocity uncertainties of $\sim$1 and $\sim$4 km s$^{-1}$ for the \fe\ and \ha\ lines, respectively. The Helioseismic and Magnetic Imager (HMI; \citealt{HMI2012}) on board the Solar Dynamics Observatory (SDO; \citealt{SDO2012}) provides the pseudo-continuum images at \fe\ $\lambda$6173 with a cadence of 45 s for the entire flare phase. We also measure the relative enhancement of this continuum. The Atmospheric Imaging Assembly (AIA; \citealt{AIA2012}) on SDO provides the EUV images at 131 \AA, in which the hot flare loops are clearly visible. The Gamma-ray Burst Monitor (GBM; \citealt{Meegan2009}) on Fermi provides HXR data in the energy range from 8 keV to 40 MeV. We select the data from detector N4 to invert the nonthermal electron beam parameters assuming a thick-target model \citep{Brown1971}. The Expanded Owens Valley Solar Array (EOVSA; \citealt{Gary2018}) provides microwave spectral imaging observations at 1--18 GHz. The X-Ray Sensor (XRS; \citealt{XRS1996}) on Geostationary Operational Environmental Satellite (GOES) provides the SXR emission at 1--8 \AA.


\section{Results}
\label{sec:result}

\subsection{Overview of the X1.0 white-light flare}

\begin{figure}[htb]
	\centering
	\includegraphics[width=0.85\textwidth]{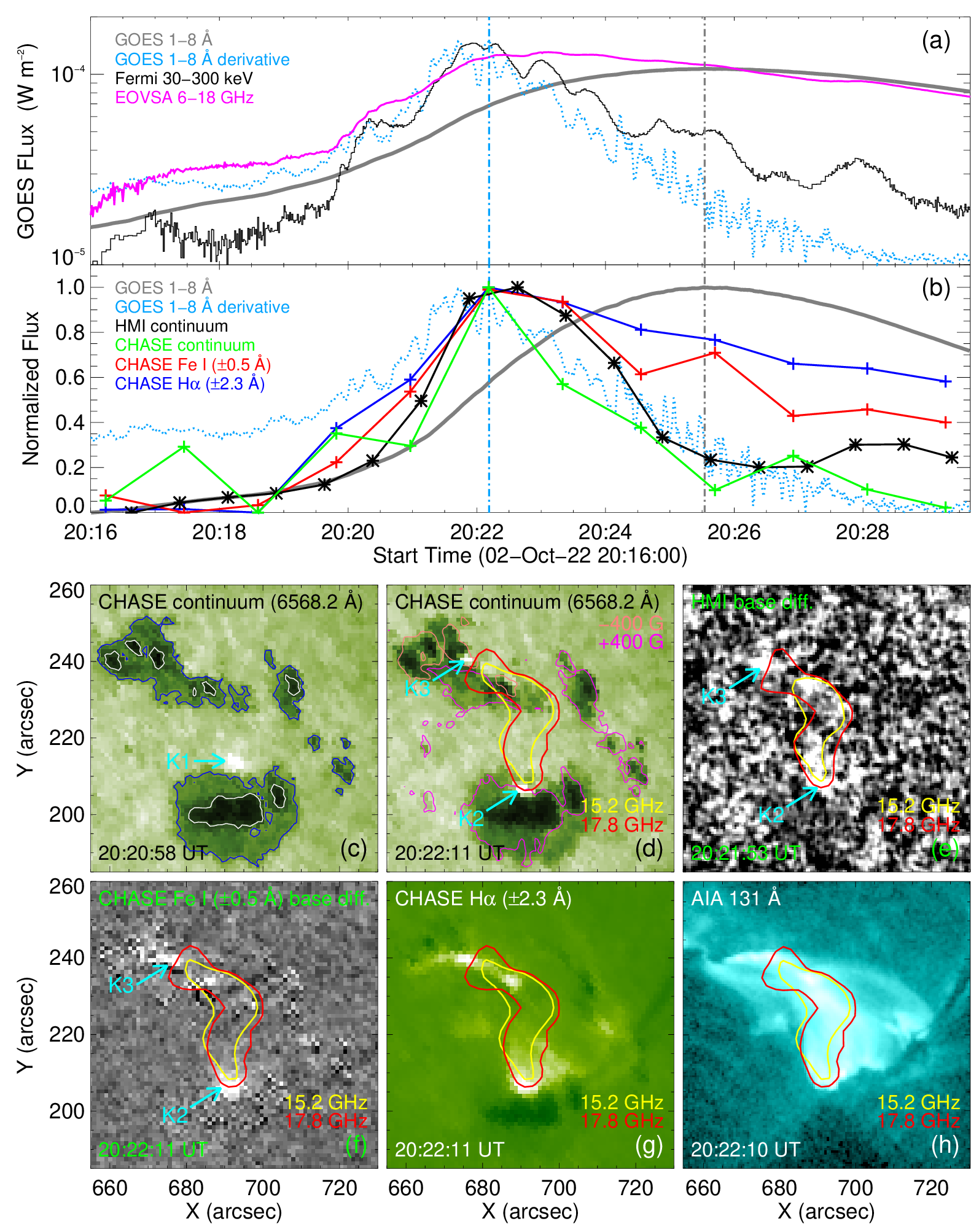}
	\caption{(a) Light curves of GOES SXR 1--8 \AA, its time derivative, Fermi HXR 30--300 keV, and EOVSA microwave 6--18 GHz for the flare. (b) HMI continuum (near 6173 \AA), CHASE continuum (at 6568.2 \AA), \fe\ (integrated over $\pm$0.5 \AA), and \ha\ (integrated over $\pm$2.3 \AA) intensity curves summed over the field of view of the images as shown in panels (c)--(g), together with the SXR 1--8 \AA\ flux and its time derivative. The gray and blue vertical lines in panels (a) and (b) denote the peak times of the SXR flux and its time derivative, respectively. (c)--(h) CHASE continuum, HMI continuum (base difference), CHASE \fe\ (base difference), CHASE \ha, and AIA 131 \AA\ images for the flare. Three white-light brightening kernels (K1--K3) are indicated by three arrows in panels (c)--(f). The white and blue contours in panel (c) mark the sunspot umbra and penumbra, respectively. The magenta and orange contours in panel (d) denote the magnetic polarities at $+$400 and $-$400 G, respectively. The yellow and red contours in panels (d)--(h) show the microwave sources at 15.2 and 17.8 GHz, respectively.}
	\label{fig1}
\end{figure}

The X1.0 flare started at 19:53 UT and peaked at 20:25 UT on 2022 October 2. From Figure \ref{fig1}(a) one can see that the Fermi HXR emission above 30 keV peaks at nearly the same time ($\sim$20:22 UT) as the time derivative of SXR flux, i.e., implying the validity of Neupert effect \citep{Neupert1968}. From Figure \ref{fig1}(b) it is seen that the HMI continuum, CHASE continuum, and \fe\ and \ha\ line intensities (integrated over the flaring region as shown in Figures \ref{fig1}(c)--(g)) also peak at $\sim$20:22 UT. Moreover, the former three appear to match the HXR emission or the time derivative of SXR flux temporally. From the CHASE (or HMI) continuum and \fe\ images (Figures \ref{fig1}(c)--(f)) we identify multiple white-light brightening kernels, denoted as K1--K3, which lie within the \ha\ ribbons (see Figure \ref{fig1}(g)). Note that the white-light kernels and \ha\ ribbons show an outward motion as the flare evolves. Kernel K1 is located outside the sunspot and shows up earlier, while K2 and K3 are in the penumbra with opposite magnetic polarities (see Figure \ref{fig1}(d)) and appear at a later time of $\sim$20:22 UT. In particular, kernels K2 and K3 seem to be conjugate footpoints of some flare loops as seen in AIA 131 \AA\ image (Figure \ref{fig1}(h)). The EOVSA microwave sources (marked by the contours in Figures \ref{fig1}(d)--(h)) also match the flare loops and their footpoints quite well. It should be mentioned that there presents a nonthermal source at the footpoints as revealed by the microwave spectral fitting. From the HXR and also microwave observations, we can conclude that there is a qualitative relationship between the white-light brightenings and the nonthermal electron beam heating. 

\subsection{Spectral features of the white-light kernels from CHASE}

\begin{figure}[htb]
	\centering
	\includegraphics[width=1.0\textwidth]{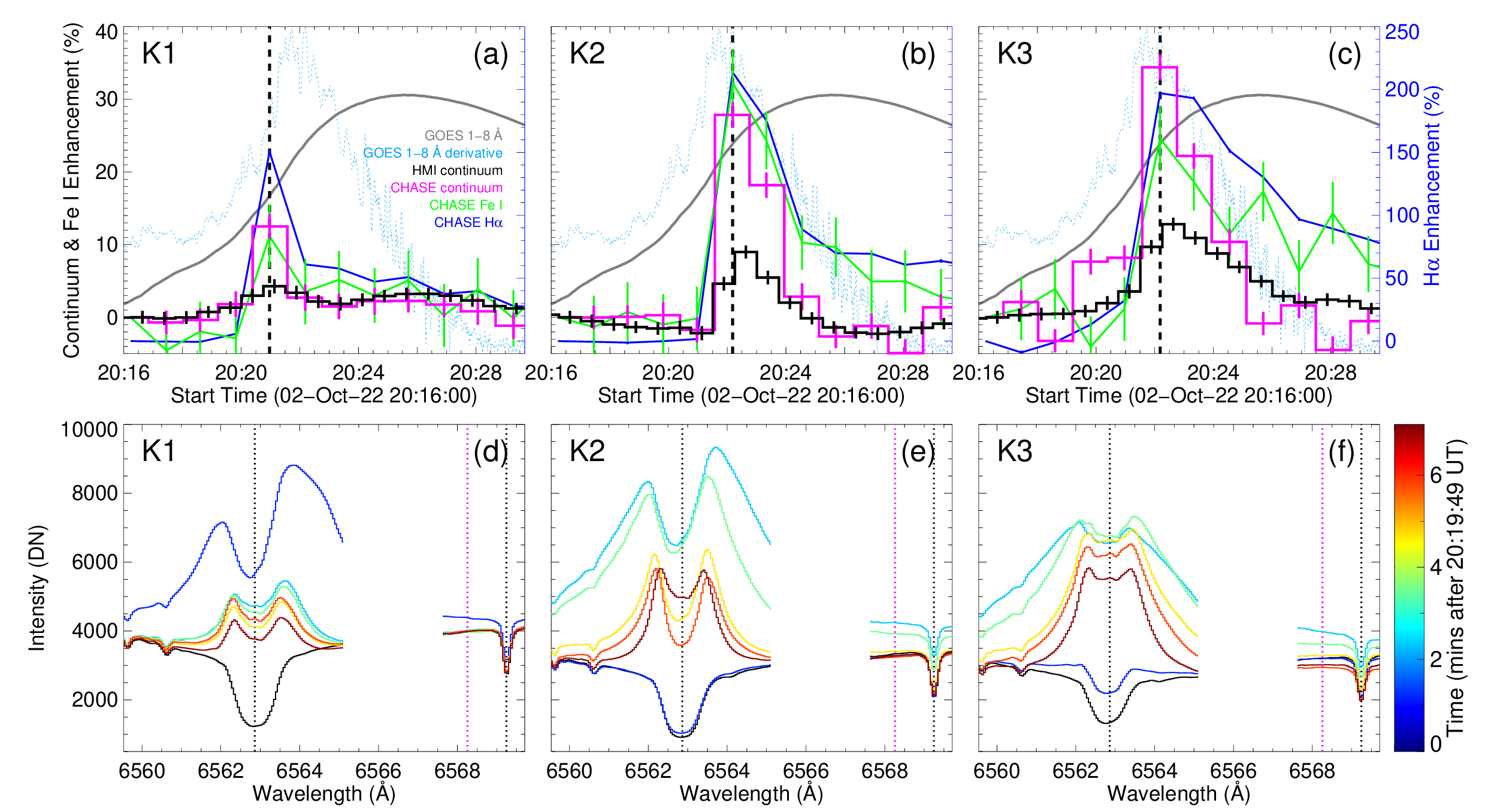}
	\caption{(a)--(c) Temporal evolution of the relative enhancements of HMI continuum, CHASE continuum, \fe\ line, and \ha\ line for kernels K1--K3 summed over an area of 3\arcsec$\times$3\arcsec, together with the SXR 1--8 \AA\ flux and its time derivative. The uncertainties of continuum and line intensity enhancements are plotted in error bars. Note that the uncertainty in \ha\ is quite small compared with its enhancement. The continuum intensity is from the wavelength at 6568.2 \AA\ as indicated by the magenta vertical line in panels (d)--(f) and the \fe\ and \ha\ line intensities are integrated over the wavelengths of $\pm$0.5 \AA\ and $\pm$2.3 \AA, respectively. The black dashed vertical line denotes the peak time of the three intensity curves from CHASE. (d)--(f) Temporal evolution of the \fe\ and \ha\ line profiles (averaged over an area of 3\arcsec$\times$3\arcsec) at K1--K3. The line centers of \fe\ and \ha\ are marked by two black dotted vertical lines in each panel.}
	\label{fig2}
\end{figure}

Figures \ref{fig2}(a)--(c) show the temporal evolution of the relative enhancements of HMI continuum, CHASE continuum, \fe\ line, and \ha\ line for kernels K1--K3 (summed over an area of 3\arcsec$\times$3\arcsec), together with the SXR flux and its time derivative. One can see that for all the kernels, the intensity curves all exhibit a sharp rise followed by a relatively gradual decay. The brightening at K1 peaks about one minute before but that at K2 and K3 peaks at nearly the same time as the time derivative of SXR flux. At K1, the maximum enhancements of HMI and CHASE continua are 4.3\% and 13\%, respectively, and those of \fe\ (integrated over $\pm$0.5 \AA) and \ha\ (integrated over $\pm$2.3 \AA) lines are 11\% and 151\%, respectively. For K2 and K3, the maximum enhancements are relatively large and up to 13\%, 34\%, 32\%, and 214\% for the four wavebands, respectively.

Figures \ref{fig2}(d)--(f) display the temporal evolution of the \fe\ and \ha\ line profiles for K1--K3 (averaged over an area of 3\arcsec$\times$3\arcsec). It is seen that at K1, the \fe\ profile has a symmetric Gaussian shape that remains in absorption during the whole period. The nearby continuum exhibits a notable enhancement at 20:21 UT (see the profile in blue in Figure \ref{fig2}(d)), so does the \fe\ line center. It should be mentioned that the blue-wing intensity of the \fe\ line rises notably at that time, which does not represent the pure continuum emission but is likely affected by the red-wing emission of the \ha\ line that is largely broadened. The \fe\ profiles and the nearby continuum at K2 and K3 are similar to those at K1. The only difference is that the relative enhancements at K2 and K3 are larger than that at K1. Finally, the \fe\ profiles at K1--K3 exhibit a very small Doppler velocity within 2 km s$^{-1}$ during the flare. For the \ha\ profiles at K1--K3, they show a significant enhancement in intensity and also a notable change in shape from absorption to emission with, however, a central reversal. The \ha\ profiles become very broad, whose wings are significantly out of the CHASE spectral window. It is interesting that the \ha\ profiles exhibit different asymmetries at the three kernels especially when the continuum is significantly enhanced. At K1 and K2, the most enhanced \ha\ profiles mainly show a red symmetry with the line center slightly blueshifted (Figures \ref{fig2}(d) and (e)). However, the most enhanced profile at K3 displays a blue asymmetry or a blue-wing enhancement (Figure \ref{fig2}(f)).

\begin{figure}[htb]
	\centering
	\includegraphics[width=1.0\textwidth]{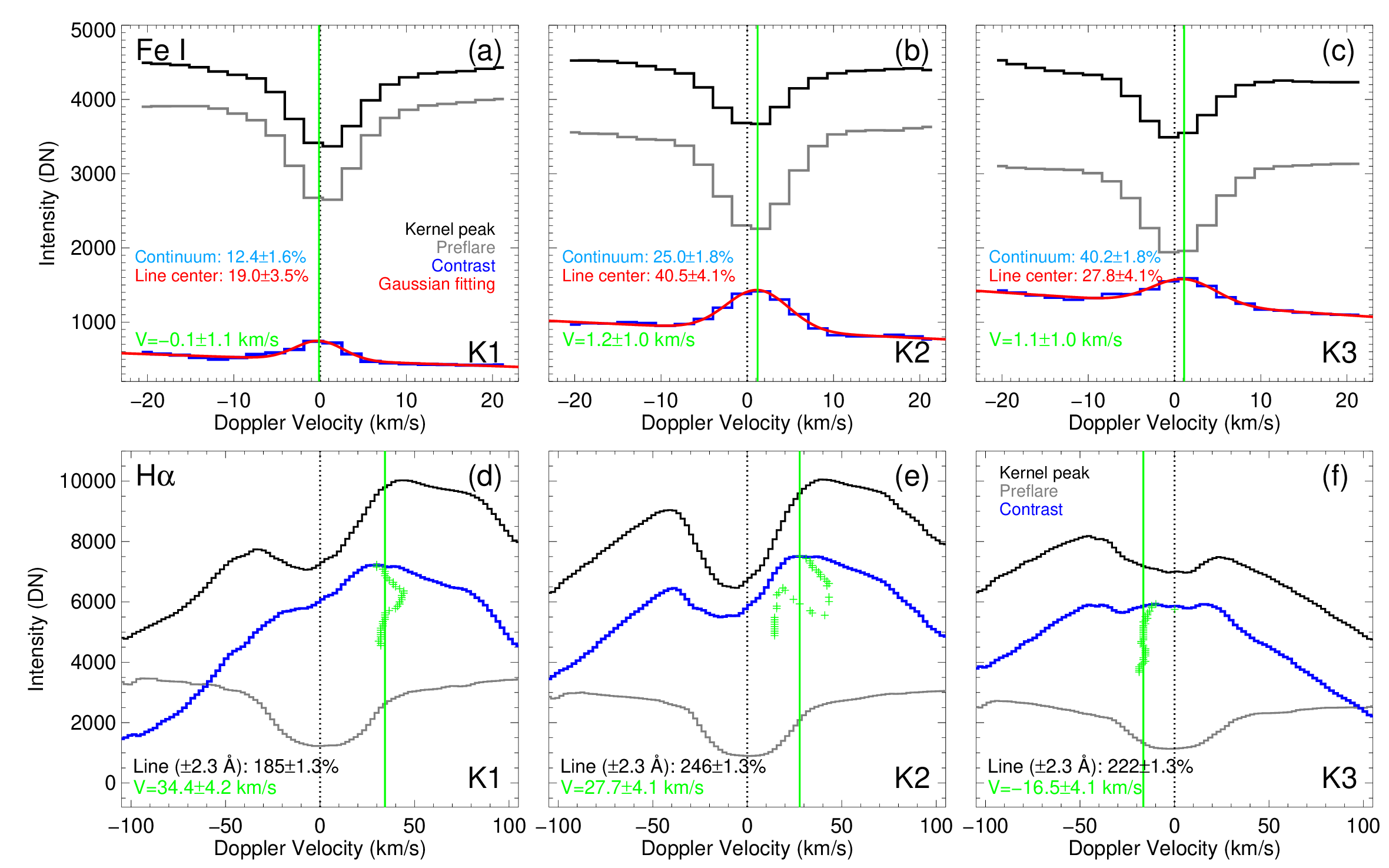}
	\caption{The line profiles of \fe\ (a--c) and \ha\ (d--f) selected from a single pixel at kernels K1--K3. The gray and black curves refer to the preflare profile and the profile at the peak time, respectively, while the blue curve represents the contrast profile (difference between them). A single Gaussian fitting and a bisector method are used to derive the Doppler velocities for \fe\ and \ha, respectively, based on the contrast profiles. The Doppler velocity plus an uncertainty is shown in each panel, together with the relative enhancements of the continuum and \fe\ and \ha\ lines. Note that for \ha, the velocity value represents the median of the velocities at different intensity levels from bisector. The black and green vertical lines in each panel indicate the line center and the median Doppler velocity, respectively.}
	\label{fig3}
\end{figure}

We further make a spectral analysis of the contrast profiles of \fe\ and \ha\ at the peak time for K1--K3, as shown in Figure \ref{fig3}. Note that the profiles here are only from a single pixel for each kernel in order not to lose spectral features due to spatial averaging. The contrast profile is defined as the one with the preflare profile subtracted. From Figures \ref{fig3}(a)--(c) we can see that all the contrast profiles of \fe\ are symmetric, which can be well fitted by a single Gaussian function. At K1, the \fe\ line center shows an enhancement of 19.0\% and the nearby continuum is enhanced with a relatively lower value of 12.4\%. The Doppler velocity of the \fe\ line at K1 is almost zero. At K2 and K3, the enhancement of \fe\ line center can be up to 40.5\% and the nearby continuum is enhanced up to 40.2\%. Note that the maximum enhancement of \fe\ line center and that of the continuum are not cospatial. The contrast profiles of \fe\ at K2 and K3 are redshifted with velocities of $\sim$1 km s$^{-1}$. For \ha\ as shown in Figures \ref{fig3}(d)--(f)), the enhancements at K1--K3 are 185\%, 246\%, and 222\%, respectively. The \ha\ contrast profiles at K1 and K2 display a notable red asymmetry and the median redshift velocities derived from bisector are 34.4 and 27.7 km s$^{-1}$, respectively. By comparison, the contrast profile at K3 exhibits a blue-wing enhancement resulting in a blueshift velocity of 16.5 km s$^{-1}$. 

\subsection{Radiative hydrodynamic simulation and comparison with observations}

To understand the white-light enhancement of this flare, we perform a radiative hydrodynamic simulation via RADYN \citep[e.g.,][]{Carlsson1992,Carlsson2002} with an electron beam heating constrained by the Fermi HXR observations. Since RADYN is a one-dimensional loop model, here we construct 18 sequential flare loops (L1--L18) corresponding to the bumps in the time derivative of SXR flux during 20:19--20:25 UT, i.e., the main phase of the flare (see Figure \ref{fig4}(a)), in a similar way to that in \cite{Rubio2016}. The inputs of each loop including the energy flux, low-energy cutoff, and spectral index of the nonthermal electron beam (as plotted in Figure \ref{fig4}(b) and also listed in Table \ref{tab1}) are obtained from the HXR spectral fitting. Two examples of the spectral fitting are shown in Figures \ref{fig4}(c) and (d) with the time intervals corresponding to the heating episodes of L6 and L11 in principle, respectively. Note that the energy flux is estimated by the total energy divided by the area of \ha\ sources that are believed to be sensitive to nonthermal electrons \citep[e.g.,][]{Fang1993}. Considering that the white-light kernels are located in the region outside the sunspot (RoS; e.g., K1) or in the penumbra (P; K2 and K3), we adopt two different model atmospheres, the quiet-Sun model VAL-C \citep{Vernazza1981} and the penumbral model \citep{Ding1989}, as the initial atmospheres. The \ha\ line is calculated in RADYN under the non-LTE condition at an emergent angle of $\mu$=0.77, while the \fe\ $\lambda$6569 and $\lambda$6173 lines are calculated via the radiative transfer code of RH \citep{Uitenbroek2001,Pereira2015} under the LTE condition \citep{hong22} with $\mu$=0.656 based on the flare location. The enhancements of the \fe\ line center and nearby continuum intensities from simulations are given in Table \ref{tab1}, which can be compared with observations for multiple white-light brightening kernels including K1--K3.

\begin{table}[h!]
\caption{Electron beam parameters in RADYN simulations and corresponding enhancements of the white-light continua and \fe\ line}
\footnotesize
\begin{center}
\scalebox{0.665}{
\makebox[\textwidth]{
\begin{tabular}{lccccccccccc}
\hline
\hline
Loop & Heating & Energy &  Low-energy & Spectral & \multicolumn{7}{c}{Relative enhancement (\%)} \\  \cline{6-12}
~ & time & flux & cutoff & index & \multicolumn{3}{c}{RADYN simulations (VAL-C/penumbral)} & ~ & \multicolumn{3}{c}{HMI \& CHASE observations} \\ \cline{6-8}  \cline{10-12}
~ & (s) & (erg s$^{-1}$ cm$^{-2}$) & (keV) & ~ & Continuum & Continuum & Line center & ~ & Continuum & Continuum & Line center \\
~ & ~ & ~ & ~ & ~ & (near 6173 \AA) & (near 6569 \AA) & (\fe\ $\lambda$6569)  & ~ & (near 6173 \AA) & (near 6569 \AA) & (\fe\ $\lambda$6569) \\ 
\hline
L1 & 11.0  & 1.9$\times$10$^{10}$  & 28.1$\pm$3.2  & 5.4$\pm$0.22 & 1.2/3.2 & 2.0/5.7 & 4.0/5.2 & ~ & ~ & ~ \\ 
\hline
\multirow{2}{*}{L2} & \multirow{2}{*}{19.0}  & \multirow{2}{*}{1.6$\times$10$^{10}$}  & \multirow{2}{*}{38.3$\pm$1.6}  & \multirow{2}{*}{6.4$\pm$0.29} & \multirow{2}{*}{ 2.2/5.3 } & \multirow{2}{*}{ 3.3/8.8 } & \multirow{2}{*}{ 4.9/6.3 } & ~ & ~ & 6.3$\pm$1.7 (RoS) & 9.4$\pm$4.1 (RoS) \\ 
~ & ~ & ~ & ~ & ~ & ~ & ~ & ~ & ~ & ~ & 6.5$\pm$1.7 (RoS) & 7.9$\pm$4.1 (RoS) \\ 
\hline
L3 & 12.0  & 2.1$\times$10$^{10}$  & 41.2$\pm$1.3  & 5.9$\pm$0.18 & 2.2/6.2 &  3.9/11.3 & 6.3/9.5   & ~ & ~ & ~ \\ 
\hline
L4 & 28.5  & 2.7$\times$10$^{10}$  & 40.6$\pm$0.9  & 4.9$\pm$0.04 & 3.1/8.2 &  4.9/13.5 & 7.6/13.7   & ~ & 8.2$\pm$0.9 (RoS) & ~ \\ 
\hline
L5 & 14.5  & 1.4$\times$10$^{10}$  & 39.5$\pm$1.4  & 4.6$\pm$0.04 & 1.8/4.6  &  2.9/8.0 & 5.1/9.9   & ~ & ~ & ~ \\ 
\hline
L6 & 17.0  & 1.3$\times$10$^{10}$  & 39.6$\pm$1.3  & 4.5$\pm$0.03 & 1.9/5.0 &  3.2/8.8 & 5.4/8.9  & ~ & 9.8$\pm$1.0 (RoS, K1) & 12.4$\pm$1.6 (RoS, K1) & 19.0$\pm$3.5 (RoS, K1) \\ 
\hline
L7 & 12.0  & 1.3$\times$10$^{10}$  & 47.2$\pm$1.1  & 4.3$\pm$0.02 & 2.0/5.6 &  3.9/11.2 & 7.4/12.4   & ~ & ~ & ~ \\ 
\hline
L8 & 13.5  & 1.4$\times$10$^{10}$  & 50.4$\pm$1.0  & 4.1$\pm$0.02 & 2.4/6.6 &  4.7/13.2 & 9.0/15.4   & ~ & ~ & ~ \\ 
\hline
L9 & 14.5  & 1.2$\times$10$^{10}$  & 72.1$\pm$1.1  & 4.2$\pm$0.02 & 2.9/7.4 &  5.6/14.5 & 11.3/18.4   & ~ & ~ & ~ \\ 
\hline
L10 & 11.7  & 1.1$\times$10$^{10}$  & 87.2$\pm$1.1  & 4.3$\pm$0.02 & 2.1/5.3 &  4.2/10.8 & 9.6/17.5   & ~ & ~ & ~ \\ 
\hline
\multirow{2}{*}{L11} & \multirow{2}{*}{19.3}  & \multirow{2}{*}{8.6$\times$10$^{9}$}  & \multirow{2}{*}{93.5$\pm$1.3}  & \multirow{2}{*}{4.4$\pm$0.02} & \multirow{2}{*}{ 2.3/5.5 } & \multirow{2}{*}{3.6/8.7} & \multirow{2}{*}{7.9/14.2} & ~ & 15.3$\pm$0.9 (P, K2) & 25.0$\pm$1.8 (P, K2) & 40.5$\pm$4.1 (P, K2) \\ 
~ & ~ & ~ & ~ & ~ & ~ & ~ & ~ & ~ & 17.6$\pm$0.9 (P, K3) & 40.2$\pm$1.8 (P, K3) & 27.8$\pm$4.1 (P, K3) \\ 
\hline
\multirow{2}{*}{L12} & \multirow{2}{*}{23.0}  & \multirow{2}{*}{4.8$\times$10$^{9}$}  & \multirow{2}{*}{96.3$\pm$1.7}  & \multirow{2}{*}{4.2$\pm$0.03} & \multirow{2}{*}{ 1.2/2.6 } & \multirow{2}{*}{1.6/3.5} & \multirow{2}{*}{5.0/9.9} & ~ & 14.5$\pm$0.9 (P) & ~ & ~ \\
~ & ~ & ~ & ~ & ~ & ~ & ~ & ~ & ~ & 21.0$\pm$0.9 (P) & ~ & ~ \\ 
\hline
L13 & 10.0  & 4.8$\times$10$^{9}$  & 94.0$\pm$1.9  & 4.3$\pm$0.03 & 0.3/1.2 &  1.1/3.1 & 4.3/7.6   & ~ & ~ & ~ \\ 
\hline
L14 & 11.0  & 5.9$\times$10$^{9}$  & 105.9$\pm$1.8 & 4.4$\pm$0.04 & 0.7/2.0 &  1.6/4.1 & 5.1/8.8   & ~ & ~ & ~ \\ 
\hline
\multirow{2}{*}{L15} & \multirow{2}{*}{24.5}  & \multirow{2}{*}{4.6$\times$10$^{9}$}  & \multirow{2}{*}{98.3$\pm$2.4}  & \multirow{2}{*}{4.2$\pm$0.04} & \multirow{2}{*}{ 1.1/2.3 }  & \multirow{2}{*}{1.5/3.2} & \multirow{2}{*}{4.2/8.1} & ~ & 13.4$\pm$0.9 (P) & 7.9$\pm$1.6 (P) & 19.5$\pm$3.8 (P) \\ 
~ & ~ & ~ & ~ & ~ & ~ & ~ & ~ & ~ & 14.2$\pm$0.9 (P) & 19.0$\pm$1.6 (P) & 25.4$\pm$3.8 (P) \\ 
\hline
L16 & 17.5  & 4.8$\times$10$^{9}$  & 77.7$\pm$2.3  & 4.1$\pm$0.04 & 0.9/2.2 &  1.8/4.6 & 5.2/7.4  & ~ & ~ & ~ \\ 
\hline
L17 & 20.0  & 5.3$\times$10$^{9}$  & 77.3$\pm$2.9  & 4.1$\pm$0.04 & 1.0/2.3 &  1.6/3.7 & 4.4/8.4  & ~ & ~ & ~ \\ 
\hline
L18 & 21.0  & 4.1$\times$10$^{9}$  & 59.4$\pm$4.6  & 3.8$\pm$0.05 & 0.9/2.2 &  1.8/4.5 & 4.5/5.9 & ~ & ~ & ~ \\ 
\hline		
\hline
\end{tabular}
}}
~~~~~~~~~~~~~~~~~~~
\end{center}
{\bf Note.} The nonthermal electron beam parameters including the energy flux, low-energy cutoff, and spectral index are time-dependent. The values here represent those at the middle time of the heating episode for each loop in RADYN simulations. The enhancements of the continua and \fe\ line from simulations represent the maximum values in each heating episode. `RoS' and `P' in observations mean that the white-light brightening kernels are located in the region outside the sunspot (corresponding to the VAL-C atmosphere) and in the penumbra, respectively.
\label{tab1}
\end{table}

The synthetic \fe\ $\lambda$6569 and \ha\ line profiles for three flare loops (L6, L11, and L4) are shown in Figure \ref{fig5}. The heating episodes of L6 and L11 correspond to the peak times of the white-light emissions from kernels K1 and K2/K3, respectively. Note that the synthetic \fe\ $\lambda$6173 profiles look very similar to the ones of \fe\ $\lambda$6569 and are not shown here. It is seen that the synthetic \fe\ and \ha\ line profiles are basically similar to the observed ones at K1--K3. The \fe\ line shows an absorption profile with the line center enhanced during the heating (Figures \ref{fig5}(a)--(c)). The nearby continuum is also enhanced due to the heating. The \ha\ line changes from absorption to emission, the latter of which shows a central reversal plus an obvious red asymmetry (Figures \ref{fig5}(d)--(f)). However, it is also seen that the continuum enhancements are only 3.2\% and 8.7\% for L6 and L11, respectively (Figures \ref{fig5}(a) and (b)), which are much lower than the observed ones of 12.4\% and 40.2\% at K1 and K3, respectively. Furthermore, the enhancements of \fe\ line center for L6 and L11 are 5.4\% and 14.2\%, respectively, which are also much lower than the observed values of 19.0\% and 40.5\% at K1 and K2, respectively. Even for L4 with the strongest heating rate and the longest heating time in our simulation (Figure \ref{fig5}(c)), the calculated enhancements are still far from sufficient to explain the observations. Similarly, the synthetic continuum near 6173 \AA\ has a much lower enhancement than the HMI observations for multiple kernels as well (see Table \ref{tab1}). It should be mentioned that for the \ha\ line, the simulations can yield a comparable enhancement with the observations for some kernels. The above results reveal that the nonthermal electron beam heating alone is not enough to produce the white-light enhancements for this flare.


\section{Summary and discussions}
\label{sec:summary}

In this Letter, we present an X1.0 white-light flare well observed in the \fe\ $\lambda$6569 and \ha\ lines by the recently launched spacecraft CHASE. This flare shows enhancements of $\sim$20\% and $\sim$40\% at the HMI and CHASE continua, respectively. The CHASE \fe\ line is also enhanced up to $\sim$40\%. At the white-light kernels, the \fe\ line shows a symmetric Gaussian shape with a small redshift of $\sim$1 km s$^{-1}$. The \ha\ line changes from absorption to emission with flare heating. The emission profiles display a central reversal and mostly exhibit a red asymmetry with the line center slightly blueshifted. There are also some \ha\ profiles showing a blue-wing enhancement. The white-light brightenings are co-spatial with the microwave footpoint sources and temporally related to the HXR emission above 30 keV. These facts imply that nonthermal electron beams play a role in contributing to the white-light enhancement. 

\begin{figure}[htb]
	\centering
	\includegraphics[width=1.0\textwidth]{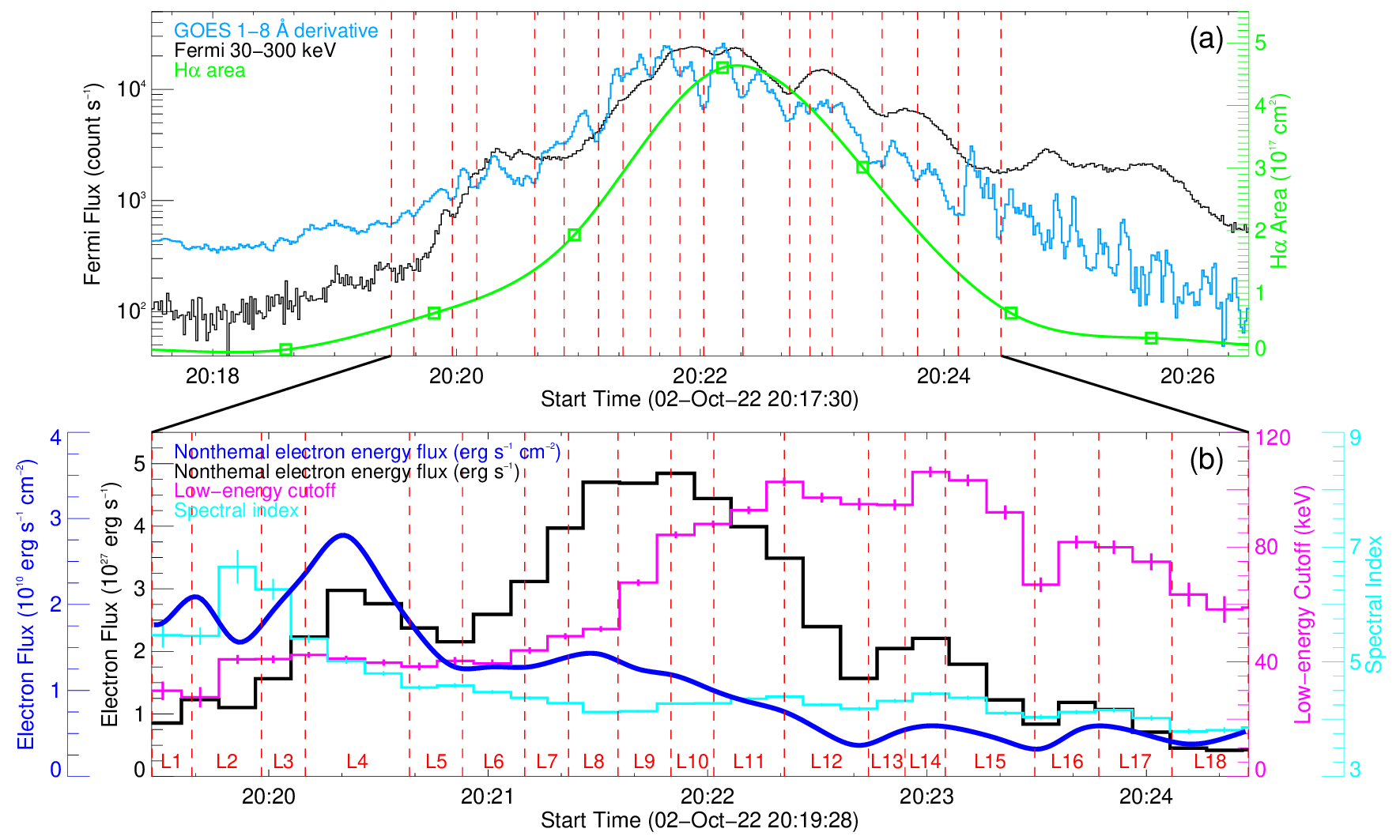}
	\includegraphics[width=0.9\textwidth]{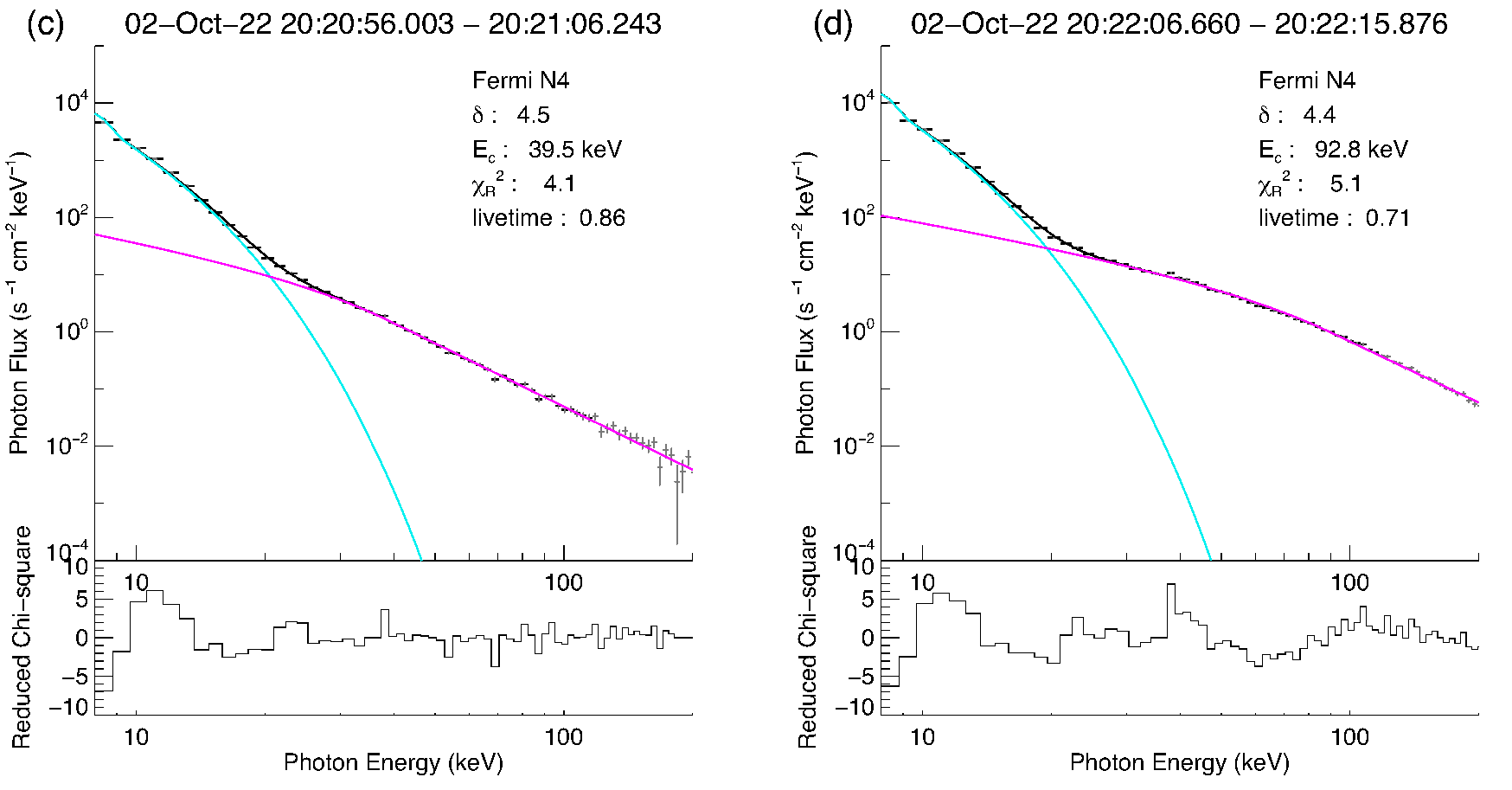}
	\caption{(a) Construction of 18 flare loops based on the GOES SXR 1--8 \AA\ time derivative plus Fermi HXR 30--300 keV emission, with the heating episodes indicated by the red vertical lines. The green curve represents the area of CHASE \ha\ brightenings after a cubic spline interpolation. (b) Electron beam parameters including the energy flux, low-energy cutoff, and spectral index used in the simulations for 18 loops (L1--L18). Note that the fitting uncertainties of low-energy cutoff and spectral index are plotted in error bars, which are quite small mostly. (c) and (d) Fermi HXR spectral fitting with a thermal component (cyan) plus a nonthermal component (magenta) for two time intervals corresponding to the heating episodes of L6 (c) and L11 (d) in principle. The fitting parameters including low-energy cutoff ($E_c$) and spectral index ($\delta$) for the nonthermal component are shown in the panels.}
	\label{fig4}
\end{figure}

\begin{figure}[htb]
	\centering
	\includegraphics[width=1.0\textwidth]{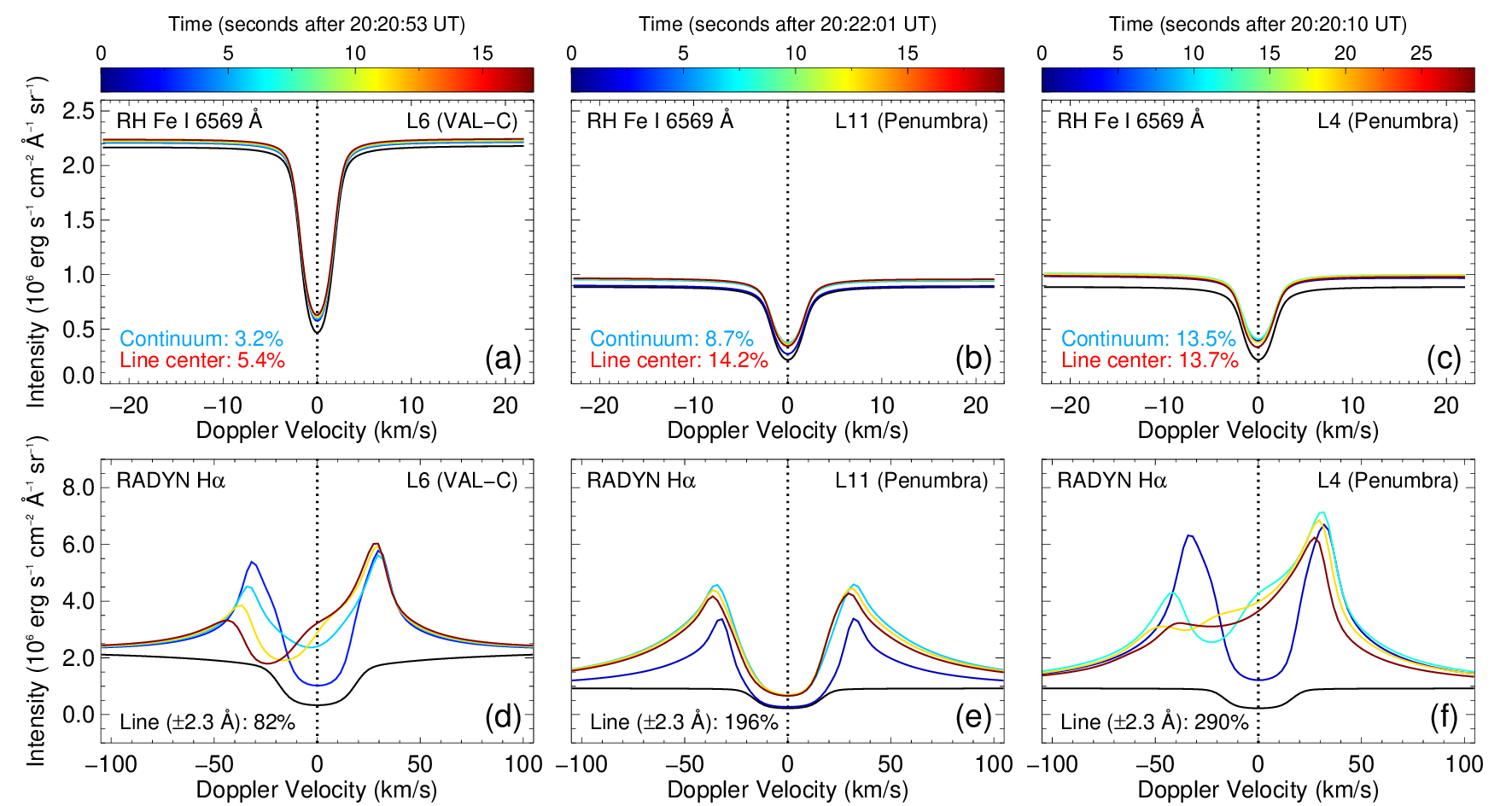}
	\caption{Synthetic \fe\ (a--c) and \ha\ (d--f) line profiles varying with time for three loops (L6, L11, and L4) in different initial atmospheres. The maximum enhancements of the continuum and \fe\ and \ha\ lines are shown for each loop from the simulations.}
	\label{fig5}
\end{figure}

In order to check if the nonthermal electron beam heating is quantitatively sufficient in producing the white-light emission, we perform observation-constrained radiative hydrodynamic simulations using RADYN. In the simulations, a key parameter is the electron energy flux, which is estimated by the total energy flux divided by the \ha\ brightening area. Doing so may suffer from a large uncertainty since the area of \ha\ brightening does not necessarily correspond to the area of electron beam heating, the latter of which could be much smaller. Therefore, we alternatively use the area of \fe\ brightening (about one forth of the \ha\ area) to obtain a higher energy flux of 3.4$\times$10$^{10}$ erg s$^{-1}$ cm$^{-2}$ in L11. This indeed generates a sufficient enhancement of the \fe\ line intensity and nearby continuum. However, this leads to an inconsistency between the calculated \ha\ line and the observed one, with the former much stronger than the latter, in particular those observed at K2 and K3. If the energy flux becomes greater than 10$^{11}$ erg s$^{-1}$ cm$^{-2}$ (for example, four times the energy flux in L4), the synthetic \fe\ profile will change from absorption to emission in the line center, which is obviously inconsistent with the CHASE observations. Another key parameter is the heating time in each loop. If increasing the heating time, we can also obtain a larger enhancement of the continuum and \fe\ line. However, a similar problem is that the \ha\ line would become too strong to match the observed one or that the \fe\ line would turn to emission incompatible with the observed line in absorption. In addition to the energy flux and heating time, the low-energy cutoff could bring an uncertainty in the synthetic white-light emission. It is known that the low-energy cutoff determined from spectral fitting is merely an upper limit in the cases where the spectral flattening caused by the cutoff is not clearly seen due to the dominant thermal emission at energies from a few to several tens of keV. This is true for some time intervals in our study (for example, the spectrum shown in Figure \ref{fig4}(c)), which may lead to an underestimation of the total energy flux of nonthermal electron beam. However, as discussed above, increasing the electron energy flux could not well reproduce the observed white-light enhancements or line profiles. In fact, the energy flux carried by the electrons with higher energies (say, $\sim$50 keV and above) that play a key role in the white-light emission \citep[e.g.,][]{Xu2006,Kuhar2016} is less affected by the uncertainty of low-energy cutoff. For the spectra with clear spectral flattening (e.g., the spectrum in Figure \ref{fig4}(d)), the cutoff energies are well determined with a relatively small uncertainty. Note that the obtained low-energy cutoff in a later time (for L9--L18) is above 50 keV, which is significantly larger than the values of 20--40 keV as usually seen in flare studies \citep[e.g.,][]{Sui2005,Aschwanden2019}. However, such high values of low-energy cutoff have also been reported \citep[e.g.,][]{Gan2002,Warmuth2009,Xia2021}. Actually, the relatively low energy electrons play a minor role (say, the backwarming effect) in generating the white-light emission. On the other hand, the uncertainties of spectral index from the HXR spectral fitting are quite small for this flare as seen in Table \ref{tab1} and Figure \ref{fig4}(b). Note that although the enhancement of the continuum near \fe\ $\lambda$6569 may be overestimated in CHASE observations due to the broadening of the nearby \ha\ line, the continuum near \fe\ $\lambda$6173 observed by HMI does not suffer from such a problem. According to the above results, we speculate that for this flare, although electron beam heating is an important source, it is not the sole one. Additional heating sources are required in order to quantitatively produce the observed continuum enhancement as well as the line profile features. Since the RADYN simulations have already included the effects of chromospheric condensation and the radiative backwarming self-consistently, the heating mechanism by \alfven\ waves deserves much attention and quantitative studies. It has been suggested that \alfven\ waves could deposit the energy in a lower layer such as the temperature minimum region in favor of generating the white-light emission \citep{Emslie1982,Fletcher2008}.

In CHASE observations, the relative enhancement of \fe\ line center is greater than that of the nearby continuum for kernels K1 and K2, while the case is reversed for K3. Based on our simulations, the enhanced continuum emission primarily originates from the layers in the middle to lower chromosphere, while the enhanced \fe\ line emission mainly comes from the bottom of the chromosphere. Therefore, the key factor is how deep the electron beam can directly penetrate in the flaring atmosphere, which depends on the preflare atmosphere and the electron beam parameters as discussed above. Considering that these parameters are quite different in these kernels, it is conceivable that they show different enhancements in the Fe I line and the continuum.

Finally, it is interesting that the observed \ha\ line profiles show a red (i.e., at K1 and K2) or blue (at K3) asymmetry when the continuum is enhanced. In particular, the asymmetries are opposite for the conjugate footpoints of K2 and K3. The asymmetry of line profiles is usually caused by a plasma flow at the line formation height. For the red asymmetric \ha\ profiles at K1 and K2, the line center is blueshifted, which could be explained by a chromospheric evaporation flow or an upflow in the formation height of \ha\ line center \citep[e.g.,][]{Kuridze2015}. In the mean time, the \ha\ line wing could be formed in the chromospheric condensation region with a downflow \citep[e.g.,][]{Ding1996}. As for the blue asymmetry of \ha\ profiles at K3, one possibility is that the line is still formed in the chromospheric condensation region that causes an absorption at the red wing \citep{Ding1996,Kuridze2015}. Another possibility is that the \ha\ line is formed in a cool upflow layer above the chromospheric evaporation region, which causes a blue-wing enhancement \citep{Tei2018}. As regards the opposite asymmetries of \ha\ at the conjugate footpoints of K2 and K3, they can be ascribed to different atmospheric conditions as well as different electron beam heating parameters at the two footpoints.

\acknowledgments
We thank the anonymous referee very much for the detailed and constructive suggestions that helped us improve the manuscript. The CHASE mission is supported by China National Space Administration. SDO is a mission of NASA's Living With a Star Program. The authors are supported by National Key R\&D Program of China under grants 2022YFF0503004 and 2021YFA1600504 and by NSFC under grants 12273115, 12127901, and 12233012.

\bibliographystyle{apj}

\end{document}